\documentclass{ws-ijmpb}
\usepackage{amsmath}
\usepackage{onlinecite}

\makeatletter

\newcommand\@pnumwidth{1.55em}
\newcommand\@tocrmarg{2.55em}
\newcommand\@dotsep{4.5}
\def\contentsname{Contents}
\newcommand\tableofcontents{%
    \section*{\contentsname
        \@mkboth{%
           \MakeUppercase\contentsname}{\MakeUppercase\contentsname}}%
    \@starttoc{toc}%
    }
\newcommand*\l@part[2]{%
  \ifnum \c@tocdepth >-2\relax
    \addpenalty\@secpenalty
    \addvspace{2.25em \@plus\p@}%
    \setlength\@tempdima{3em}%
    \begingroup
      \parindent \z@ \rightskip \@pnumwidth
      \parfillskip -\@pnumwidth
      {\leavevmode
       \large \bfseries #1\hfil \hb@xt@\@pnumwidth{\hss #2}}\par
       \nobreak
       \if@compatibility
         \global\@nobreaktrue
         \everypar{\global\@nobreakfalse\everypar{}}%
      \fi
    \endgroup
  \fi}
\newcommand*\l@section[2]{%
  \ifnum \c@tocdepth >\z@
    \addpenalty\@secpenalty
    \addvspace{1.0em \@plus\p@}%
    \setlength\@tempdima{1.5em}%
    \begingroup
      \parindent \z@ \rightskip \@pnumwidth
      \parfillskip -\@pnumwidth
      \leavevmode \bfseries
      \advance\leftskip\@tempdima
      \hskip -\leftskip
      #1\nobreak\hfil \nobreak\hb@xt@\@pnumwidth{\hss #2}\par
    \endgroup
  \fi}
\newcommand*\l@subsection{\@dottedtocline{2}{1.5em}{2.3em}}
\newcommand*\l@subsubsection{\@dottedtocline{3}{3.8em}{3.2em}}
\newcommand*\l@paragraph{\@dottedtocline{4}{7.0em}{4.1em}}
\newcommand*\l@subparagraph{\@dottedtocline{5}{10em}{5em}}

\let\citeonline\citelow

\makeatother

\numberwithin{equation}{section}

\newcommand{\bp}{{\bf p}}
\newcommand{\bk}{{\bf k}}
\newcommand{\bK}{{\bf K}}
\newcommand{\br}{{\bf r}}
\newcommand{\bR}{{\bf R}}
\newcommand{\kF}{k_{\mathrm{F}}}
\newcommand{\kB}{k_{\mathrm{B}}}

\begin{document}

\markboth{N. H. March, G. G. N. Angilella, and R. Pucci}
{Natural orbitals in relation to quantum information theory}

%
\catchline{}{}{}{}{}
%

\title{\uppercase{Natural orbitals in relation to quantum information theory:
from model light atoms through to emergent metallic properties}}

\author{N. H. MARCH}

\address{Department of Physics, University of Antwerp,\\
Groenenborgerlaan, 171, B-2020 Antwerp, Belgium\\[.5\baselineskip]
Oxford University, Oxford, UK}

\author{G. G. N. ANGILELLA\footnote{Corresponding author. E-mail:
giuseppe.angilella@ct.infn.it}}

\address{Dipartimento di Fisica e Astronomia, Universit\`a di Catania,\\
Via S. Sofia, 64, I-95123 Catania, Italy\\[.5\baselineskip]
Scuola Superiore di Catania, Universit\`a di Catania,\\ Via
Valdisavoia, 9, I-95123 Catania, Italy\\[.5\baselineskip]
CNISM, UdR Catania, Via S. Sofia, 64, I-95123 Catania, Italy\\[.5\baselineskip]
INFN, Sez. Catania, Via S. Sofia, 64, I-95123 Catania, Italy}

\author{R. PUCCI}

\address{Dipartimento di Fisica e Astronomia, Universit\`a di Catania,\\
Via S. Sofia, 64, I-95123 Catania, Italy\\[.5\baselineskip]
CNISM, UdR Catania, Via S. Sofia, 64, I-95123 Catania, Italy}

\maketitle

\begin{history}
\received{\today}
\end{history}

\begin{abstract}
The review begins with a consideration of 3 forms of quantum information entropy
associated with Shannon and Jaynes. For model two-electron spin compensated
systems, some analytic progress is first reported. The Jaynes entropy is clearly
related to correlation kinetic energy. A way of testing the usefulness of a
known uncertainty principle inequality is proposed for a whole class of model
two-electron atoms with harmonic confinement but variable electron-electron
interaction.

Emerging properties are then studied by reference to bcc Na at ambient pressure
and its modeling by `jellium'. Jellium itself has collective behaviour with
changes of the density, especially noteworthy being the discontinuity of the
momentum distribution at the Fermi surface. This has almost reduced to zero at
$r_s = 100$~a.u., the neighbourhood in which the quantal Wigner electron solid
transition is known to occur. However, various workers have studied crystalline
Na under pressure and their results are compared and contrasted. Work by DFT on
K, Rb, and Cs is discussed, but now with reduced density from the ambient
pressure value. The crystalline results for the cohesive energy of these metals
as a function of lattice parameters and local coordination number are shown to
be closely reproduced by means of ground and excited states for dimer potential
energy curves.

Then, pair potentials for liquid Na and Be are reviewed, and compared with the
results of computer simulations from the experimental structure factor for Na.

Finally, magnetic field effects are discussed. First a phenomenological model of
the metal-to-insulator transition is presented with an order parameter which is
the discontinuity in the Fermi momentum distribution. Lastly, experiments on a
two-dimensional electron assembly in a GaAs/AlGaAs heterojunction in a
perpendicular magnetic field are briefly reviewed and then interpreted.
\end{abstract}



\tableofcontents

\markboth{N. H. March, G. G. N. Angilella, and R. Pucci}
{Natural orbitals in relation to quantum information theory}

\clearpage

\section{Background and outline}

The starting point of the present review is the work of Amovilli and March
\cite{Amovilli:04b}. These authors drew attention to the importance of natural
orbitals (NOs) \cite{Loewdin:55a} in quantum information theory. These NOs will
be defined in Section~\ref{sec:definition} below, together with a brief summary
of some of their properties. But they are still somewhat difficult to obtain,
and even in light atoms like He and Be, one must then have recourse to numerical
methods. Therefore, in Section~\ref{sec:models}, known explicit analytic forms
of NOs for light model atoms will be used to gain insight.
Section~\ref{sec:entropy} follows Amovilli and March \cite{Amovilli:04b} in
expressing quantum information entropy in terms of NO properties.

We then turn for the remainder of this review to emergent properties, with
emphasis placed on metallic states. 
We shall
essentially deal with periodic crystalline Na and Be, with the former modelled,
at atmospheric pressure, by the Sommerfeld, or as now commonly described
`jellium', model. Thus, in metallic bcc Na, the monovalent ionic charges are
smeared out into a uniform background of positive charges, neutralizing the
interacting electron assembly. A considerable merit of this jellium model is
that the NOs are plane waves $e^{i\bk\cdot\br}$: an enormous simplification.
However, other properties, including the ground-state energy per electron, are
still only known numerically from quantum Monte~Carlo simulations, starting with
the pioneering work of Ceperley and Alder \cite{Ceperley:80,Senatore:94}. In
particular, by now, the occupation numbers (or the momentum distribution) of the
plane wave states are known as a function of variable electron density $\rho_0 =
3/(4\pi r_s^3)$, with $r_s$ the mean interelectronic spacing.

Turning to both Na and Be metals without smearing the ions, we note in
Section~\ref{sec:periodic} for the benefit of the reader a proof that the NOs
have Bloch wave form, \emph{viz.} $e^{i\bk\cdot\br} u_\bk (\br)$, where $u_\bk
(\br)$ is a periodic function with the period of the lattice.
Section~\ref{sec:alkalipressure} then contains a discussion of the remarkable
effect of applying high pressure to the alkali metals Li, Na, and K.
Section~\ref{sec:Beliq} treats Be liquid metal in some detail. A
phenomenological model of a metal-to-insulator transition in then summarized in
Section~\ref{sec:MI}. Because of key experiments using an GaAs/AlGaAs
heterojunction in a magnetic field, a summary of such experiments plus some
theoretical work is given in Section~\ref{sec:2djellium}. The final section
constitutes a summary, plus some proposals for future research, both theoretical
and experimental, that should be fruitful.

\section{Definition of natural orbitals (NOs) and calculation for a model of
He-like atomic ions}
\label{sec:definition}

Following L\"owdin \cite{Loewdin:55a}, we first use the exact (normalized)
ground state wave function $\Psi$ for an $N$-electron assembly to define a
so-called first-order density matrix $\gamma(\br^\prime , \br)$ by
\begin{equation}
\gamma(\br^\prime , \br) = N \int \Psi^\ast (\br^\prime , \br_2 , \ldots \br_N )
\Psi (\br , \br_2 , \ldots \br_N ) d\br_2 \cdots d\br_N .
\label{eq:1DM}
\end{equation}
This leads to the ground-state density $\rho(\br)$, the central tool of DFT
\cite{Parr:89}, as
\begin{equation}
\rho(\br) = \left. \gamma(\br^\prime , \br) \right|_{\br^\prime = \br} .
\end{equation}
But L\"owdin's proposal was to define (normalized) natural orbitals $\phi_i
(\br)$ and their occupation numbers $n_i$ which bring $\gamma(\br^\prime , \br)$
into the diagonal form
\begin{equation}
\gamma(\br^\prime , \br) = \sum_i n_i \phi_i^\ast (\br^\prime ) \phi_i (\br) .
\end{equation}
The remainder of this Section will give
illustrative examples, for admittedly simplistic models, of the occupation
numbers $n_i$ and the orbitals $\phi_i (\br)$.

\subsection{Explicit forms of NOs for light model atoms, and construction of the
first-order density matrix (1DM)}
\label{sec:models}

As an example of the (albeit approximate) calculation of NOs and their
occupation numbers, we shall consider, following Amovilli, Howard, and March
(AHM) \cite{Howard:05b,Amovilli:05,Amovilli:08} a change in the Hamiltonian $H_{\mathrm{T}}$
proposed long ago by Temkin \cite{Temkin:62} (see also
Ref.~\citeonline{Poet:78}). $H_{\mathrm{T}}$ was obtained from the exact
non-relativistic Hamiltonian for He-like atomic ions by replacing the
electron-electron interaction $u(r_{12}) = e^2 /r_{12}$, where $r_{12} = | \br_1
-\br_2 |$ denotes the relative coordinate of the two electrons, by its spherical
average, which becomes then solely $u(r_1 , r_2)$, with no longer dependence on
the angle between $\br_1$ and $\br_2$. The Hamiltonian $H_{\mathrm{AHM}}$ was
then taken to be defined by
\begin{equation}
H_{\mathrm{AHM}} = H_{\mathrm{T}} + \delta (r_1 - r_2),
\label{eq:AHM}
\end{equation}
where the $\delta$ function represents an additional radial correlation in an
admittedly simplistic manner. AHM \cite{Howard:05b,Amovilli:05,Amovilli:08}
solve this Hamiltonian exactly for non-relativistic He positive ions with external
potential
\begin{equation}
V_{\mathrm{ext}} (\br) = -\frac{Ze^2}{r} ,
\end{equation}
this being, of course, already present in $H_{\mathrm T}$. Here, we shall focus
on their study of the NOs for this Hamiltonian, Eq.~(\ref{eq:AHM}). If $\Psi$
denotes the symmetric ground-state wave function, the NOs diagonalize the
density matrix kernel operator
\begin{equation}
\gamma(r,r^\prime ) = 4\pi \int_0^\infty d r_2 \, r_2^2 \Psi(r,r_2 ) \Psi(r_2 ,
r^\prime ) .
\label{eq:kernop}
\end{equation}
Therefore, by spectral decomposition, we can write
\begin{equation}
\gamma(r,r^\prime ) = \sum_j n_j \phi_j (r) \phi_j (r^\prime ) ,
\label{eq:spectraldecomposition}
\end{equation}
where $\sum_j n_j = 1$. Since we have just two electrons, the form
(\ref{eq:kernop}) allows one to write
\begin{equation}
\Psi (r_1 , r_2 ) = \sum_j \nu_j \phi_j (r_1 ) \phi_j (r_2 ) ,
\end{equation}
with $\nu_j^2 = n_j$ and
\begin{equation}
4\pi \int_0^\infty d r_2 \, r_2^2 \Psi (r_1 , r_2 ) \phi_j (r_2 ) = \nu_j \phi_j
(r_1 ).
\label{eq:eigen}
\end{equation}
As Amovilli and March (AM) \cite{Amovilli:05} show, the eigenvalue sum $\sum_j
\nu_j$ can also be found. Taking the wave function at coincidence ($r_1 = r_2$),
after integration one finds
\begin{eqnarray}
\sum_j \nu_j &=& \frac{8\pi N}{(2Z-1)^3} \nonumber\\
&=& \frac{8(Z-1)}{2Z-1} \sqrt{\frac{2Z^2 -3Z +1}{32 Z^3 -50 Z +20}} \nonumber\\
&=& 1 - \frac{15}{32} \frac{1}{Z} + O \left( \frac{1}{Z^2} \right) ,
\end{eqnarray}
the last expansion applying in the limit of large $Z$. AM \cite{Amovilli:05}
then obtain a set of approximate NOs. The first 10 occupation numbers and the
corresponding eigenvalues $\nu_j$ as obtained by AM are reproduced in
Table~\ref{tab:AMnuj}.

\begin{table}[t]
\tbl{First 10 occupation numbers and eigenvalues of the spectral
decomposition of the ground-state wave function for He in the modified Temkin
$s$-wave model. After Ref.~\protect\citelow{Amovilli:05}.}{%
\begin{tabular}{rr@{.}lr@{.}l}
\Hline\\[-1.8ex] 
$j$ & \multicolumn{2}{c}{$\nu_j$} & \multicolumn{2}{c}{$n_j$} \\[0.8ex]
\hline \\[-1.8ex]
1 &   0  & 97856 & 0 & 95758 \\
2 & $-0$ & 19988 & 0 & 03995 \\
3 & $-0$ & 04333 & 0 & 00188 \\
4 & $-0$ & 01911 & 0 & 00037 \\
5 & $-0$ & 01079 & 0 & 00012 \\
6 & $-0$ & 00694 & 0 & 00005 \\
7 & $-0$ & 00484 & 0 & 00002 \\
8 & $-0$ & 00358 & 0 & 00001 \\
9 & $-0$ & 00275 & 0 & 00001 \\
10& $-0$ & 00218 & 0 & 00000 \\[0.8ex]
\Hline \\[-1.8ex]
\end{tabular}
}
\label{tab:AMnuj}
\end{table}

\subsection{Differential equation for NOs of AHM Hamiltonian}
\label{ssec:odeAHM}

The integral equation~(\ref{eq:eigen}) can, as AM have shown \cite{Amovilli:05},
be transformed into a second-order linear differential equation satisfied by the
NOs $\phi_i$. Thus AM find first the integral equation
\begin{equation}
e^{r_1} \int_0^{r_1} dr_2 \, r_2^2 e^{-Zr_2} \phi_j (r_2 ) + \int_{r_1}^\infty
dr_2 \, r_2^2 e^{-(Z-1)r_2} \phi_j (r_2 ) = \frac{\nu_j}{4\pi N} e^{Zr_1} \phi_j
(r_1 ) .
\label{eq:AMB1}
\end{equation}
Dividing through $e^{r_1}$ and differentiating again, the above equation reduces
to a homogeneous second-order linear differential equation for $\phi_j$. After
some modest manipulation, AM show that the result reads
\begin{equation}
\phi_j^{\prime\prime} + (2Z-1) \phi_j^\prime + \left( Z(Z-1) - \frac{4\pi N
r^2}{\nu_j} e^{-(2Z-1)r} \right) \phi_j = 0.
\label{eq:AMB3}
\end{equation}
This equation is what must be hoped is a forerunner of a corresponding equation
of realistic rather than model He-like ions. Eq.~(\ref{eq:AMB3}) possesses
solutions for any value of $\nu_j \in [-1,1]$ apart from $\nu_j =0$, with
behaviour at small and large $r$ as listed below
\begin{subequations}
\begin{eqnarray}
\phi_j &\to& \exp (-Zr) , \quad r\to0,\\
\phi_j &\to& \exp (-(Z-1)r) , \quad r\to\infty.
\end{eqnarray}
\end{subequations}
For further details on $\nu_j$ and $n_j$, the interested reader must consult
Ref.~\citelow{Amovilli:05}.

A proposed characterization of the correlated 1DM for the ground-state of the
Hookean model for a two-electron atom can be found in
Ref.~\citeonline{March:12a}.

\section{Quantum information entropy and NO properties}
\label{sec:entropy}

In most books on DFT, formulae are written for quantum entropy $S$, in suitable
units, in terms of the ground-state density $\rho(\br)$, $S_\rho$ say, or
alternatively the corresponding momentum density $n(\bp)$, say $S_n$. These read
\begin{subequations}
\begin{align}
S_\rho &= \mathrm{const} \times \int \rho(\br) \ln \rho(\br) d \br \\
\intertext{and}
S_n &= \mathrm{const} \times \int n(\bp) \ln n(\bp) d \bp .
\end{align}
\label{eq:Shannon}
\end{subequations}
Of course, to be truly interesting, $S_\rho$ and $S_n$ should be calculated from
fully correlated ground-state densities in both $\br$ and $\bp$ space. Such
densities are now obviously available in numerical form only, however, using well-known quantum-chemical
techniques like the coupled-cluster method.

In fact, as Amovilli and March \cite{Amovilli:04b} pointed out, a quantity known
also to information theory workers, the so-called Jaynes entropy, and related to
the correlated 1DM $\gamma(\br^\prime ,\br)$, could be evaluated solely in terms
of occupation numbers $n_i$ introduced in Section~\ref{sec:definition} above.
The L\"owdin natural orbitals $\phi_i (\br)$ are not required, the result for
an $N$-electron system being \cite{Amovilli:04b}
\begin{equation}
S_{\mathrm{Jaynes}} = \mathrm{const} \times \sum_i \frac{n_i}{N} \ln \left( \frac{n_i}{N} \right) .
\label{eq:Jaynes}
\end{equation}

As a more recent example than those cited in Section~\ref{sec:models}, it is
relevant here to mention the model studied by Schilling \emph{et al.}
\cite{Schilling:13}. For spinless Fermions in one dimension, these authors
studied a generalized Moshinsky Hamiltonian (see Ref.~\citeonline{March:08a} and
references therein), namely
\begin{equation}
H = \sum_{i=1}^N \left( \frac{p_i^2}{2m} + \frac{1}{2} m\omega^2 x_i^2 \right) +
\frac{1}{2} D \sum_{i,j=1}^N (x_i - x_j )^2 .
\end{equation}
Natural lengths they single out are identified by
\begin{subequations}
\begin{align}
\ell &= \sqrt{\frac{\hbar}{m\omega}} ,\\
\tilde{\ell} &= \sqrt{\frac{\hbar}{m\omega\sqrt{1+ND/m\omega^2}}} ,
\end{align}
\end{subequations}
for $N$ identical Fermions. They in fact parametrize the coupling by
writing
\begin{equation}
\delta = \ln \left( \frac{\ell}{\tilde{\ell}} \right) = \frac{1}{4} \ln \left(
1 + \frac{ND}{m\omega^2} \right) .
\end{equation}
They obtain then the exact correlated 1DM as
\begin{equation}
\gamma(x^\prime ,x) = p(x^\prime ,x) \exp \left( -\alpha (x^{\prime2} +
x^2 ) + \beta x^\prime x \right) ,
\end{equation}
where $p$ denotes a symmetric polynomial of degree $2(N-1)$ in the variables
$x^\prime$ and $x$, while $\alpha$ and $\beta$ are constants depending on
$\ell$, $\tilde{\ell}$, and $N$.

To summarize their achievement concerning the occupation numbers, they note a
duality property
\begin{equation}
n_i (\delta ) = n_i (-\delta)
\end{equation}
relating attractive ($\delta<0$) and repulsive ($\delta>0$) Fermion-Fermion
interactions. This means that expansions of $n_i$ in powers of $\delta$ contain
only even order terms. Schilling \emph{et al.}
\cite{Schilling:13} give the first eight occupation numbers to $O(\delta^{10})$,
for the case $N=3$. So, at least in this admittedly simplistic example, the
Jaynes entropy via Eq.~(\ref{eq:Jaynes}) can be estimated.

\section{Emergent properties, especially to metallic states}
\label{sec:emergent}

\subsection{NOs of periodic crystal: proof that they have Bloch wave form}
\label{sec:periodic}

As we consider crystalline properties, it is of obvious interest to enquire what
is the nature of the NOs in such assemblies (see also
Ref.~\citeonline{Jones:73-1}). Let us expand the ground-state wave function
$\Psi$ in products of single-particle orthonormal functions as
\begin{equation}
\Psi = \sum_{\ell_1 ,\ldots \ell_N} c(\ell_1 ,\ldots \ell_N ) \phi_{\ell_1}
(\br_1) \cdots \phi_{\ell_N} (\br_N) ,
\end{equation}
where $N$ is the number of electrons. Note that the coefficients must be chosen
to take care of Fermion antisymmetry. We may write for the 1DM that
\begin{equation}
\gamma(\br^\prime , \br) = \sum_{k\ell} \alpha_{k\ell} \phi_k^\ast (\br^\prime)
\phi_\ell (\br) ,
\label{eq:gammaJM}
\end{equation}
where
\begin{equation}
\alpha_{k\ell} = N \sum_{\ell_1 ,\ldots \ell_N} c^\ast (k,\ell_2 ,\ldots \ell_N)
c (\ell,\ell_2 ,\ldots \ell_N ) .
\end{equation}
Since the $\alpha_{k\ell}$ form an Hermitean matrix, it can therefore be brought
into diagonal form. Let $b_{\ell m}$ the elements of the diagonalizing matrix. Next
define the orthonormal set
\begin{equation}
\psi_\ell = \sum_m b_{\ell m} \phi_m .
\end{equation}
Eq.~(\ref{eq:gammaJM}) then becomes
\begin{equation}
\gamma(\br^\prime , \br) = \sum_{\ell} a_\ell \psi_\ell^\ast (\br^\prime )
\psi_\ell (\br) ,
\label{eq:all}
\end{equation}
as discussed earlier in this Review. The orbitals $\psi_\ell$ are the NOs and
the $a_\ell$ are occupation numbers.

Here, it is to be noted that, if $\Psi$ is normalized to unity, then
\begin{subequations}
\begin{align}
\sum_{\ell_1 ,\ldots \ell_N} | c(\ell_1 , \ldots \ell_N ) |^2 &= 1 \\
\intertext{and}
\sum_i a_i &= N . 
\end{align}
\end{subequations}
It is further to be noted that the coefficients $c(\ell_1 , \ldots \ell_N )$ are
antisymmetric in $\ell_1 , \ldots \ell_N$, since $\Psi$ is an antisymmetric wave
function. This implies that one can write
\begin{equation}
\Psi = \sum_K c_K \Psi_K ,
\end{equation}
where $K$ denotes a `configuration' $\ell_1 < \ell_2 < \ldots \ell_N$, while
$\Psi_K$ is a normalized determinant
\begin{equation}
\Psi_K = \left(\frac{1}{N!}\right)^{1/2} \det \psi_i (\br_j) .
\label{eq:pseudoHF}
\end{equation}
It is readily shown that $c_K = (N!)^{1/2} c(\ell_1 , \ldots \ell_N )$ amd
\begin{equation}
\sum_K | c_K |^2 = 1.
\label{eq:cK1}
\end{equation}
In the Hartree-Fock (HF) approximation, $\Psi$ is a single determinant having
the form (\ref{eq:pseudoHF}) above, an the $a_\ell$ of Eq.~(\ref{eq:all}) are
either zero or unity. Generalizing this, it can be shown that
\begin{equation}
a_\ell = \sum_{K(\ell)} | c_K |^2 ,
\end{equation}
the sum extending over all configurations including $\ell$. But because of
Eq.~(\ref{eq:cK1}), it follows that
\begin{equation}
0 \leq a_\ell \leq 1 \qquad \mbox{all $\ell$.}
\end{equation}
This summarizes the so-called Pauli conditions on the 1DM.

This is the stage to show that, for an ideal crystal, the NOs must have Bloch
function form. That is, they are such that
\begin{equation}
\psi_\bk (\br + \bR_\mu ) = e^{i\bk\cdot\bR_\mu} \psi_\bk (\br)  ,
\label{eq:Bloch}
\end{equation}
for all vectors $\bR_\mu$ in the Bravais lattice. To prove this, it is to be
noted that, taking periodic boundary conditions,
\begin{equation}
|\Psi (\br_1 + \bR_\mu , \ldots \br_N + \bR_\mu )|^2 = 
|\Psi (\br_1 , \ldots \br_N )|^2 ,
\end{equation}
which merely expresses the physical equivalence of every cell in a perfect
crystal. From the above equation, it follows that
\begin{equation}
\Psi (\br_1 + \bR_\mu , \ldots \br_N + \bR_\mu )
=
\Psi (\br_1 , \ldots \br_N ) \exp (i\bK\cdot\bR_\mu ).
\end{equation}
Next one can expand in orthogonal Bloch functions $\phi_{\ell_i \bk}$, where
$\ell_i$ denotes an index corresponding to the band index of the independent
particle model:
\begin{equation}
\Psi = \sum_{\mathrm{all}\,\ell\bk} c(\ell_1 \bk_1, \ldots \ell_N 
\bk_N ) \phi_{\ell_1 \bk_1} (\br_1 ) \cdots \phi_{\ell_N \bk_N} (\br_N ).
\end{equation}
It is easy to show that the sum over the $\bk$'s is restricted by the condition
\begin{equation}
\sum_{i=1}^N \bk_i = \bK ,
\end{equation}
so that one may write
\begin{equation}
\gamma(\br,\br^\prime) = \sum_{\ell m \bk} \alpha_{\ell m} (\bk)
\phi_{\ell\bk}^\ast (\br^\prime) \phi_{m\bk} (\br) .
\end{equation}
Thus, for a set of Bloch wave functions, the occupation numbers are always
diagonal in $\bk$.

It remains to diagonalize in the `band' indices $\ell$ and $m$. The
diagonalizing matrix $b_{\ell m}$ is dependent on $\bk$, but being unitary, the
NOs
\begin{equation}
\psi_{\ell\bk} (\br) = \sum_m b_{\ell m} (\bk) \phi_{m\bk} (\br)
\end{equation}
obey condition (\ref{eq:Bloch}) since the $\phi_{m\bk}$ do so. As a special
case, the NOs of homogeneous jellium must be plane waves: a fact we have already
utilized in an earlier section.

\section{Li, Na, and K under pressure}
\label{sec:alkalipressure}

\subsection{Alkali metals under extreme conditions, especially Li and Na}
\label{ssec:LiNa}

Because of their relatively simple electronic structure and relatively high
electron density, light alkali metals in normal conditions, including Li, Na, and
K, have long been considered to have a `simple' metallic behaviour \cite{AM}.
The free-electron model was also believed to work even better at higher pressure
and electron density. Such a `tenet' was first reconsidered by Siringo \emph{et
al.} \cite{Siringo:97}, who, already in the title of their work, explicitly
posed the question whether the light alkali metals are still metals under high
pressure. These authors studied an extended Hubbard model on a bcc lattice,
characterized by an on-site electron repulsion $U$ and an intersite electron
interaction $V$. The main result of their work was that such a model electron
system should undergo a metal-to-insulator transition with increasing electron
density. Such a transition occurred at a critical value of the adimensional
Wigner-Seitz radius of $r_s = 2.6$, corresponding to a pressure $p=100$~GPa for
Na. Later works by the same authors even predicted re-entrant metallicity at yet
higher pressure, corresponding to oscillations between a symmetric (metallic)
phase and a low-symmetry (dimerized and insulating) phase, as a consequence of
Friedel oscillations in the electron pair potential
\cite{Angilella:02e,Angilella:03c}.

The original finding by Siringo \emph{et al.} \cite{Siringo:97} was
followed by DFT studies on dense lithium by Neaton and Ashcroft
\cite{Neaton:99}, who predicted a structural instability of bcc Li towards a
more complex Cmca structure at high pressure. In this structure, each Li ion is
coordinated by another ion only, and the pairing thereof can give rise to a
metal-to-insulator transition at some $r_s < 2.1$. According to Neaton and
Ashcroft \cite{Neaton:99}, the initial distortion is due to the Jahn-Teller
mechanism, while further distortion is due to a balance between the Peierls
instability and exchange effects.

The above theoretical findings immediately kindled much experimental interest
\cite{Hanfland:99,Hanfland:00,Fortov:99}, which confirmed the tendency of the
light alkali metals to lower their symmetry under high pressure. Low symmetry
phases have been indeed observed for Li \cite{Hanfland:00} and Na
\cite{Neaton:01,McMahon:07}. In particular for Li, Hanfland \emph{et al.}
\cite{Hanfland:00} established experimentally and theoretically that a cI16
structure stabilizes at $p\simeq 40$~GPa and $T\simeq 180$~K. The existence of a
broken symmetry phase was confirmed by the observation of Goncharov \emph{et
al.} \cite{Goncharov:05} of a Raman vibrational band in Li above 70~GPa to
120~GPa, the thus confirming the theoretical prediction by Matsuoka \emph{et
al.} \cite{Matsuoka:08} and by Struzhkin \emph{et al.} \cite{Struzhkin:02} of
the existence of two further phases of Li, \emph{viz.} Li-VI and Li-VII, which
are stable at $T=25$~K and $p=69$~GPa and $p=86$~GPa, respectively. Furthermore,
Matsuoka and Shimizu \cite{Matsuoka:09}, by performing electrical measurements
up to 105~GPa, have inequivocally found that a metal-to-insulator transition
occurs in Li near 80~GPa at $T<50$~K. Preliminary calculations
\cite{Christensen:01} seemed to indicate that both Li and Na show a tendency
towards the formation of atomic pairs, and a dimerized oC8 structure was
predicted to be the most stable phase above 165~GPa for Li, and above 220~GPa
for Na. According to more recent and compelling calculations, several
proposals have been made for the identification of the structures of the Li-VI
and Li-VII phases, including the oC88 and oC40 structures \cite{Guillaume:11},
the c2-24 and Aba2-24 structures \cite{Yao:09}, and more recently an Aba2-40
structure \cite{Marques:11,Lv:11}.

Analogies and differences between Li and Na under pressure have been described
in detail by Rousseau \emph{et al.} \cite{Rousseau:11}. Like Li, Na crystallizes
in the bcc structure at ambient conditions, and undergoes a structural
transition towards the fcc phase at 65~GPa \cite{Hanfland:02}. Increasing
pressure up to $p=103$~GPa, Na takes the Na-III phase
\cite{McMahon:07,Gregoryanz:08}, which has a cI16 symmetry, which is similar to
the symmetry adopted by compressed Li. However, above $p=117$~GPa, the Na-III
phase tranforms into the Na-IV phase \cite{Gregoryanz:08,Lundegaard:09}, which
is characterized by an oP8 symmetry. Yet two more phases of Na have been
discovered, \emph{viz.} Na-V above $p=125$~GPa
(Refs.~\citeonline{Gregoryanz:08,Lundegaard:09}), and Na-VI above $p=200$~GPa
(Ref.~\citeonline{Ma:09}). This latter phase presents a simple double-hexagonal
close-packed (dhcp) structure, with hP4 symmetry, and is typically transparent
\cite{Ma:09}. The hP4 structure has been confirmed also theoretically by Gatti
\emph{et al.} \cite{Gatti:10}.

The physical reason for dimerization is an open question. \emph{Ab initio}
electronic structure calculations \cite{Rousseau:00} indicated a tendency
towards distance alternation, due to a sizeable overlap of $p\pi$ orbitals in
the interstitial regions, and to electron localization between ions
\cite{Rousseau:08,MartinPendas:99}. On the other hand, it has been proposed
\cite{Christensen:01,Bergara:00} that the increase of $s$-$p$ hybridization
could give rise to a low coordination number. However, even the fully dimerized
phase is far from any standard covalent solid: the electron density is uniformly
spread and almost constant, while the first and second neighbours distances are
comparable. Such a dimerized phase is better described as a charge density wave
in a high density metal rather than a molecular solid. Such a low-symmetry phase
in distorted cubic-based structures, as those observed experimentally in the
compressed alkalis, has been connected by Angilella \emph{et al.}
\cite{Angilella:02e,Angilella:03c} to the presence of Friedel oscillations in
the electronic pair potential. These can ultimately be traced back to the
existence of a sharp Fermi surface, and should therefore be generic to most
metals.

The deceivingly `simple' light alkali metals are a continuous source of
experimental surprises and theoretical challenges. Recently, earlier predictions
of superconductivity in compressed lithium \cite{Christensen:01a} have been
confirmed experimentally \cite{Shimizu:02a,Struzhkin:02,Deemyad:03,Matsuoka:08}.
At variance with Li, the superconducting critical temperature $T_c$ of Na is
predicted to be rather small ($T_c < 1$~K) also in fcc phase
\cite{Christensen:06,Shi:06}. If constrained in structures with reduced
dimensionality (thin layers or wires), both lithium and sodium have been
predicted to exhibit other electronic instabilities, such as a ferromagnetic one
\cite{Bergara:03,Ashcroft:01}. Closer looks at the phase diagram of both lithium
and sodium at high pressure indicate that these light alkali metals should
rather prefer the liquid state, with lithium being thus the elemental metal with
the lowest melting point
\cite{Gregoryanz:05,Gregoryanz:08,Guillaume:11,Boehler:83,Lazicki:10,Schaeffer:12}.
Several theoretical works have also been performed both on Li and on Na
\cite{Tamblyn:08,Hernandez:07,Hernandez:10,Raty:07,Martinez-Canales:08,Koci:08}.
Further \emph{ab initio} calculations \cite{Rousseau:05,Ma:08,Pickard:09,Lv:11}
at yet higher pressure indicate more complicated crystal structures, a tendency
towards a metal-to-semiconducting transition, and the increasingly relevant role
of core orbitals in establishing the electronic behaviour of the compressed
alkalis. However, both for lithium \cite{Matsuoka:09} and sodium \cite{Ma:09} a
metal-to-insulator transition at high pressure has now been experimentally well
established.

\subsection{Mainly expanded potassium related to dimer potential energy curves}

Relatively early work on potassium subjected to high pressure was carried out by
March and Rubio \cite{March:97a}. These authors utilized DFT ground-state theory
in the local density approximation (LDA).

A variety of different crystal structures were studied in the above work, the
bcc structure at atmospheric pressure, with coordination number $z=8$, being the
natural starting point. Of course, the appropriate free energies are all fairly
close, so that the intersections showing phase transitions due to change in $z$
are, no doubt, sensitive to the approximation (LDA) chosen for the
exchange-correlation potential $V_{\mathrm{xc}} (\br)$ of DFT.

We give here a brief summary of the DFT computations of March and Rubio
\cite{March:97a}. They were concerned with the way cohesive energies depended on
local coordination number $z$ and on bond length. However, unlike the above
discussion on the light alkalis, incrasing bond length was a focus, due to
experiments of Winter \emph{et al.} \cite{Winter:88,Winter:89} on the heavier
alkalis and in particular Rb and Cs. These experiments, both neutron scattering
and thermodynamic, were along the liquid-vapour coexistence curve up to the
critical point. This was only accessible experimentally for these two alkalis.
It was striking from the neutron measurements of the structure factor $S(k)$
that $z$ decreased dramatically from a high value $\sim 8-10$ near the triple
point to about 2 at the critical point. One of us \cite{March:89a,March:90}
fitted the experimental data for density $d$ by
\begin{equation}
d=az+b
\end{equation}
where $a=0.23$~g$\cdot$cm$^{-3}$ and $b=-0.08$~g$\cdot$cm$^{-3}$ (see also
Ref.~\citeonline{Freeman:94}).

\begin{figure}[t]
\centering
\includegraphics[height=0.8\columnwidth,angle=-90]{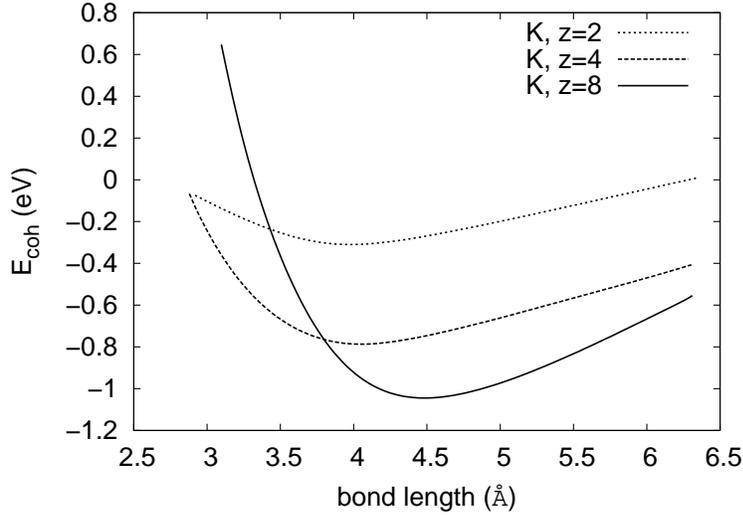}
\caption{Cohesive energy for K lattices corresponding to coordination numbers
$z=8$ (bcc), 4 (diamond), and 2 (chain) as a function of the nearest-neighbout
bond length, from \emph{ab initio} calculations. Redrawn after
Ref.~\protect\citeonline{March:97a}.}
\label{fig:Rubio1}
\end{figure}

\begin{figure}[t]
\centering
\includegraphics[height=0.8\columnwidth,angle=-90]{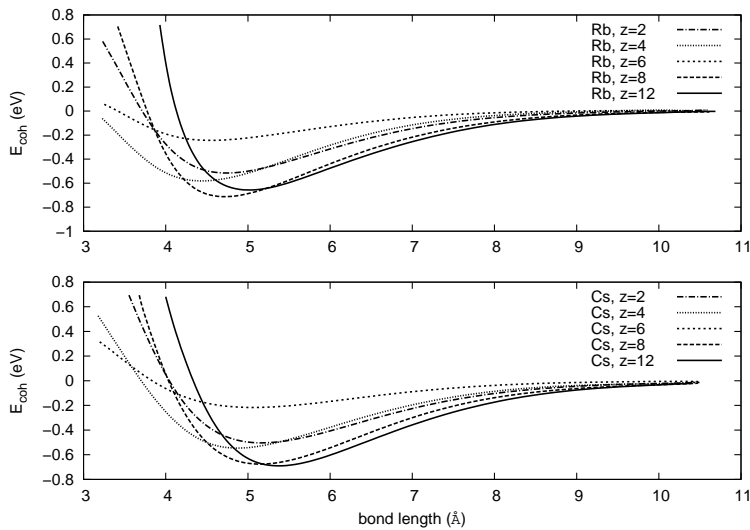}
\caption{Same as in Fig.~\protect\ref{fig:Rubio1}, but now for Rb (upper panel)
and Cs (lower panel) atoms with long-range order in bcc, diamond, fcc, simple
cubic (sc), and linear chain structures (coordinations $z=2-12$). Redrawn after
Ref.~\protect\citeonline{March:97a}.}
\label{fig:Rubio2}
\end{figure}

To return now to the calculations of March and Rubio
\cite{March:97a}, these LDA calculations by DFT were such that the cohesive
energies for mainly K, but also for Rb and Cs, were studied as a function of
bond length and coordination number. The essential theoretical predictions are
shown in Figs.~\ref{fig:Rubio1} and \ref{fig:Rubio2}. The chain structures of Rb
and Cs were studied for reasons referred above, though the relevant experiments
were on fluids, not crystals as in the March-Rubio calculations
\cite{Angilella:04k}.

With regard to our emphasis in this review on emergent properties, we note that
March and Rubio \cite{March:97a} could interpret their DFT results in terms of
dimer potential energy curves, as predicted by March, Tosi, and Klein
\cite{March:95a}. The underlying idea of these authors was to decompose the
cohesive energy $E(z,r_0 )$ into a linear combination of two terms. The first of
these is proportional to the coordination number $z$ and is to be viewed as a
typical pairwise additive contribution, which can be written in the form
$\frac{1}{2} z R(r_0)$, following March \emph{et al.} \cite{March:95a}. The
quantal chemical interpretation of $r_0$ is related to the potassium free-space
dimer potential energy curve. The second contribution will be written in the
form $-f(z)g(r_0)$, both $R$ and $g$ being related to free-space dimer
properties. As March and Rubio \cite{March:97a} emphasize, their DFT results on
expanded potassium can be represented very usefully by this simple
quantum-chemical approach.

Concerning potassium under pressure, it should be noted that, like Li and Na
discussed in Section~\ref{ssec:LiNa}, also compressed potassium is
characterized by several structural transitions among cubic-like phases, thus
confirming the not-so-simple character of the alkalis also in this compound.
Numerous works are available in the literature, both experimental and
theoretical, and the reader is especially referred to
Refs.~\citeonline{McMahon:06,Lundegaard:09a,Marques:09,Loa:11,Lundegaard:13,Alouani:89}.

\section{Emergent properties in liquid metallic Be}
\label{sec:Beliq}

We note here, first of all, the emergence of metallic Be, as we bring Be atoms,
initially widely separated, together to eventually form the equilibrium hcp
lattice, at zero temperature. The electron density in the unit cell was, as
discussed by Matthai \emph{et al.} \cite{Matthai:80}, characterized by marked
angularity. Their work \cite{Matthai:80} drew heavily on the X-ray Bragg
relection data of Brown \cite{Brown:72} as well as approximate Wannier functions
\cite{Jones:73-1} representing Bloch waves in the NOs. Of course, when Be metal
is heated sufficiently, liquid Be, an unpleasant toxic fluid, forms.

Now, when we speak of electron density, it corresponds to taking a `snapshot' of
the nuclear positions at a given instant. Since, at the melting temperature,
$T_m$ say, of Be metal, the electrons are still essentially totally degenerate
fermions, we can proceed to calculate the electron density. But to do practical
calculations on the monovalent liquid alkalis, or divalent liquid Be, it is
necessary to appeal to `neutral pseudoatoms'
\cite{Corless:61,Worster:63,Ziman:64}.

Therefore, Perrot and March \cite{Perrot:90a,Perrot:90} calculated, by DFT, the
electron density $\Delta \rho_{\mathrm{ps}} (r)$ associated with such a
pseudoatom (ps) for the conduction electrons in liquid metallic Be. Their
density $\Delta \rho_{\mathrm{ps}} (r)$ 
is presented in the useful form $Q(r)$ defined such that $Q(r)$ gives the number
of (conduction) electrons, two per atom in Be of course, within a sphere of
radius $r$ drawn with centre on the nucleus of the chosen pseudoatom. The
remarkable oscillations in this $Q(r)$ are then due to the emergent Fermi
surface in the metal, taken in the liquid, naturally enough, to be a sphere of
radius $\kF$, the Fermi wave number. The oscillations were predicted by Blandin
and Friedel \cite{Blandin:59} and independently by March and Murray
\cite{March:60a}.

One of the consequences of such a `displaced' charge density around a divalent
Be positive ion is a resultant pair potential $\phi(r)$, which, of course,
depends on the (now liquid) density of the conduction electrons, Perrot and
March \cite{Perrot:90a,Perrot:90} being concerned only with the liquid metal
near $T_m$ at ambient pressure.

\begin{figure}[t]
\centering
\includegraphics[height=0.8\columnwidth,angle=-90]{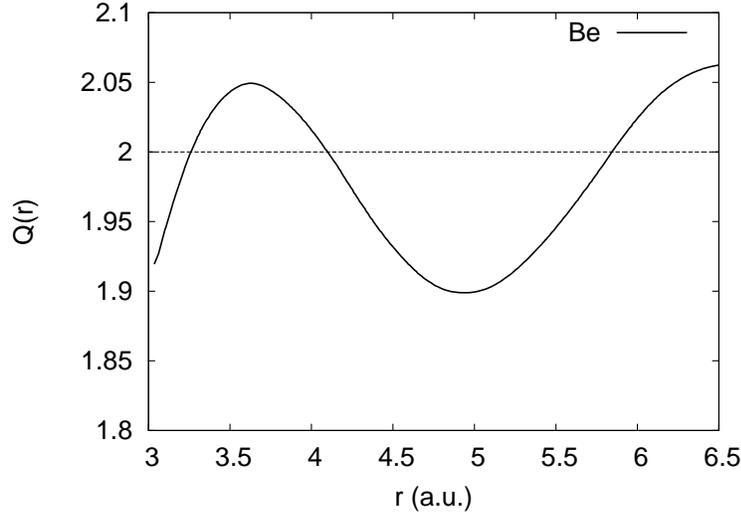}
\caption{Total valence screening charge $Q(r)$ for liquid Be at a density equal
to the solid density. After Ref.~\protect\citelow{Perrot:90}.}
\label{fig:4p314book}
\end{figure}

\begin{figure}[t]
\centering
\includegraphics[height=0.8\columnwidth,angle=-90]{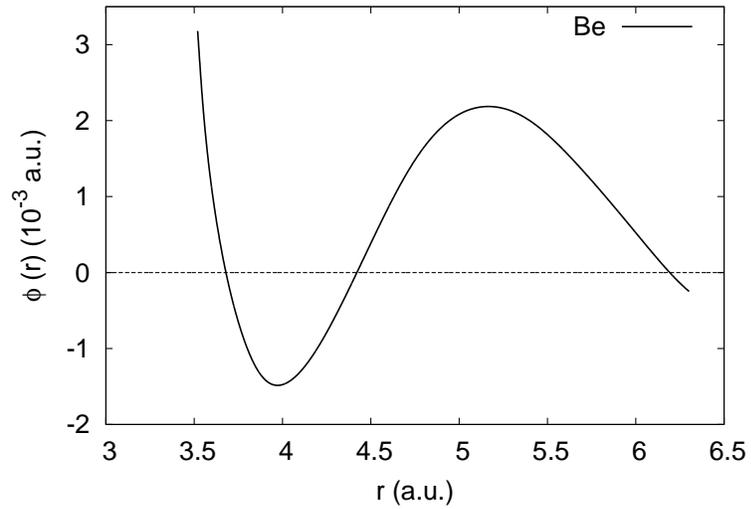}
\caption{Pair potential $\phi(r)$ for liquid Be. After Ref.~\protect\citelow{Perrot:90}.}
\label{fig:5p314book}
\end{figure}

If we write $Q(r)$ quite explicitly as
\begin{equation}
Q(r)= \int_0^r 4\pi r^2 \Delta \rho_{\mathrm{ps}} (r) dr ,
\label{eq:prePoisson}
\end{equation}
then from Poisson's equation the potential $V_e (r)$ created by the above
displaced charge is related directly to $Q(r)$ as follows:
\begin{equation}
\frac{1}{r^2} \frac{d}{dr} \left( r^2 \frac{dV_e}{dr} \right) = -4\pi \Delta \rho_{\mathrm{ps}} (r)
\end{equation}
and hence, using Eq.~(\ref{eq:prePoisson}), one readily obtains
\begin{equation}
\frac{dV_e}{dr} = - \frac{Q(r)}{e^2} .
\end{equation}
But, as discussed especially by Corless and March \cite{Corless:61},
\begin{equation}
\phi(r) = -z V(r) ,
\end{equation}
where $z$ is the valency ($z=2$, for Be metal). Hence the turning points of the
pair potential $\phi(r)$ occur at positions $r$ where
\begin{equation}
\frac{dV}{dr} = \frac{dV_e}{dr} - \frac{z}{r^2} = 0,
\end{equation}
or
\begin{equation}
Q(r) = z .
\end{equation}
Fig.~\ref{fig:4p314book} shows $Q(r)$ versus $r$, while the Perrot-March pair
potential for Be is shown in
Fig.~\ref{fig:5p314book}.

Quite recently, Hoggan and March \cite{Hoggan:12} have compared this pair
potential $\phi(r)$ between the (screened) ions in liquid Be with the potential
energy curve of the free-space dimer. The free-space dimer has a single (and
very shallow) turning point (minimum) not very far from the value $r = 4$~a.u.,
which can be read off Fig.~\ref{fig:5p314book}. This stresses, in the context of
the present review, the dramatic way the free space dimer potential energy curve
is altered in the emergent liquid metal.

\begin{figure}[t]
\centering
\includegraphics[height=0.8\columnwidth,angle=-90]{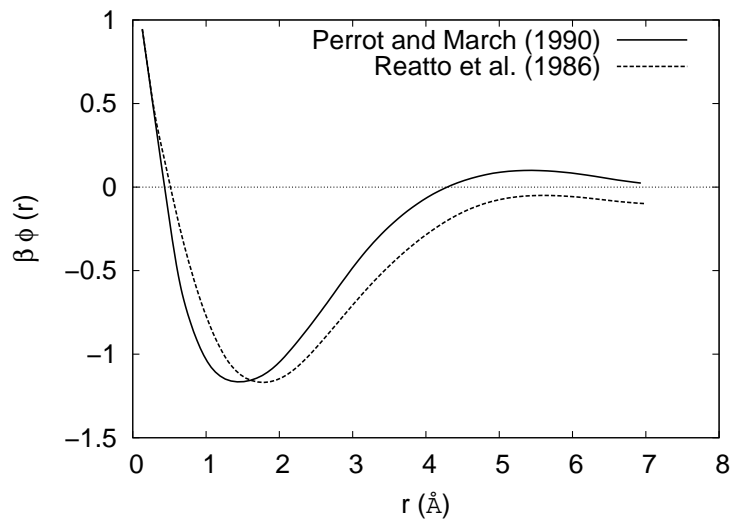}
\caption{Pair potentials for liquid Na at $T=100$~$^\circ$C and density of
0.929~g$\cdot$cm$^{-3}$. The upper curve for large $r$ is Perrot and March
\protect\cite{Perrot:90a} prediction, while lower curve for large $r$ has been derived
from inversion of liquid structure factor in the experiments by Reatto \emph{et
al.} \protect\cite{Reatto:86,Reatto:88}. After Ref.~\protect\citelow{Perrot:90a}.}
\label{fig:Reatto}
\end{figure}

While the pair potential in Fig.~\ref{fig:5p314book} for liquid Be has not been
tested to date, corresponding calculations of Perrot and March \cite{Perrot:90a}
for liquid Na can be brought into precise contact with a potential extracted
from the measeured iquid structure factor $S(k)$. This has been used by Reatto
\emph{et al.} \cite{Reatto:86,Reatto:88}, following the procedure set out by
Johnson and March \cite{Johnson:63}. Therefore, we can plot the Reatto pair
potential against the theoretical prediction of Perrot and March
\cite{Perrot:90a} for liquid Na nearing the freezing point, and the results are
shown in Fig.~\ref{fig:Reatto}. The lower curve at large $r$ is the experimental
prediction of Reatto \emph{et al.} \cite{Reatto:86,Reatto:88}, while the other
curve is the electron theory result of Perrot and March \cite{Perrot:90a}. All
the main features are seen to be accurately predicted by the electron theory.

\section{Phenomenological theory of first- and second-order metal-to-insulator
(MI) transitions at $T=0$}
\label{sec:MI}

Below, a phenomenological theory of metal-to-insulator (MI) transitions at $T=0$
will be outlined \cite{March:79}, in which the discontinuity $q$ in the
single-particle occupation probability $n(p)$ at the Fermi surface is the order
parameter. We apply it to the case of the second order MI transition in a
half-filled Hubbard band. The enhancement of the spin susceptibility by the
Hubbard interaction will also be discussed. Going back to jellium already
discussed above, the same phenomenology is applied here, but now a first order
MI transition occurs when the conduction electron liquid crystallizes onto the
Wigner lattice.

\subsection{Spin susceptibility near metal-to-insulator (MI) transition}

Then, following March \emph{et al.} \cite{March:79}, the energy expansion reads,
for non-zero magnetization $m$ and applied magnetic field $h$,
\begin{equation}
E(m,q) = R_0 + a m^2 -hm + E_1 q + E_2 q^2 + \ldots + e q m^2 .
\end{equation}
Minimizing with respect to $m$ yields
\begin{equation}
2am + 2 e q m = h,
\end{equation}
and hence the spin susceptibility $\chi$ can be read off from
\begin{equation}
m = \frac{1}{2} [a(U) + e q]^{-1} h \equiv \chi h ,
\end{equation}
where $U$ denotes the on-site repulsion within Hubbard model. If a transition
of the second order occurs at a critical value $U=U_c$, then $q\to0$ as $U\to
U_c$, and the physical question as to the enhancement of $\chi$ as the MI
transition is approached rests on the behaviour of $a(U)$ as a function of the
Hubbard interaction $U$. Provided $a\to 0$ as $1-U/U_c$ or faster, it can be
shown \cite{March:79} that
\begin{equation}
\chi \propto \left( 1 - \frac{U}{U_c} \right)^{-1} .
\end{equation}
This has been discussed also by Brinkman and Rice \cite{Brinkman:70} in relation
to experiments on V$_2$O$_3$ which indeed show enhancement of $\chi$ by $U$ in
the metallic phase as the MI is approached. 

\subsection{Magnetic susceptibility of expanded fluid alkali metals}

Related to the above phenomenology of a correlation-induced MI transition,
Chapman and March \cite{Chapman:88} have proposed an interpretation of the
behaviour of the magnetic susceptibility of expanded fluid metals.
Renormalization of the Fermi temperature caused by correlation is shown to play
a major role. An important conclusion of Chapman and March \cite{Chapman:88} is
that the momentum distribution at the Fermi surface is quantitatively very
different from jellium at the same density.

\section{Jellium in two-dimensions in an applied magnetic field}
\label{sec:2djellium}

\subsection{Magnetically induced quantal Wigner solid as produced in GaAs/AlGaAs
heterojunction}

Magnetically aided quantal Wigner electron crystals were first predicted
theoretically in 1968 by Durkan \emph{et al.} (DEM) \cite{Durkan:68}. These
authors were motivated by experiments of Putley \cite{Putley:60} on highly
compensated $n$-type InSb. Putley observed conduction in an impurity band as a
function of applied magnetic field of strength $B$, which however was found
experimentally to be suppressed at a critical magnetic field $B_c$ as $B$ was
gradually increased. DEM argued that, at this transition, Wigner quantal
electron crystallization was occurring. Subsequent experiments by Somerford
\cite{Somerford:71} on the same system were interpreted by Care and March
\cite{Care:71} as a Wigner transition aided by the magnetic field. Later,
Kleppmann and Elliott \cite{Kleppmann:75} pressed this interpretation and showed
that the anisotropy of the conductivity measured by Somerford was consistent
with the Wigner transition.

Some 20 years after the above proposal of DEM, interest in the magnetically
induced Wigner electron solid (MIWS) was revived by the beautiful experiment of
Andrei \emph{et al.} \cite{Andrei:88} on a two-dimensional electron assembly in
a GaAs/GaAlAs heterojunction. These authors pointed out that an essential
difference between an electron liquid (delocalized state) and a quantal Wigner
electron solid is that it is only the latter phase that can sustain low
frequency shear waves. Evidence was presented by Andrei \emph{et al.}
\cite{Andrei:88} that their 2D electrons, in a sufficiently strong magnetic
field perpendicular to the plane of the electron assembly, could support shear
waves.

\begin{figure}[t]
\centering
\includegraphics[height=0.8\columnwidth,angle=-90]{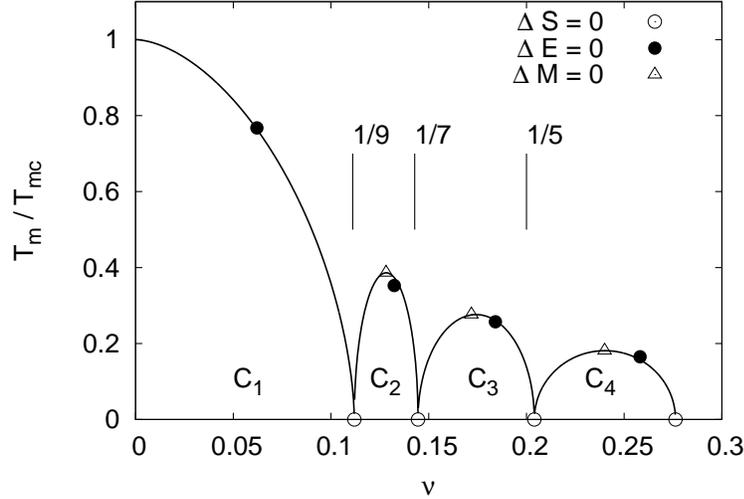}
\caption{Schematic phase diagram as proposed by Buhmann \emph{et al.}
\protect\cite{Buhmann:91}, showing the four crystal phases, $C_1$ to $C_4$, and the
reentrant liquid phase at $\nu=\frac{1}{9}$, $\frac{1}{7}$, and $\frac{1}{5}$.
Ordinate show the melting temperature $T_m$, normalized with respect to that of
a classical one-component plasma \protect\cite{Andrei:88}, $T_{mc} = e^2 (\pi n)^{1/2} /
\kappa \kB \Gamma_m$, where $\kB$ is Boltzmann's constant, $\kappa$ the
dielectric constant of the host material, and $\Gamma_m = 127 \pm 3$ (see
Ref.~\protect\citelow{Lea:91}, and references therein). Symbols mark the points $\Delta
M$, $\Delta S$, $\Delta E = 0$, as deduced from thermodynamics. Redrawn after
Ref.~\protect\citelow{Lea:91}.
}
\label{fig:1p100book}
\end{figure}

\begin{figure}[t]
\centering
\includegraphics[height=0.8\columnwidth,angle=-90]{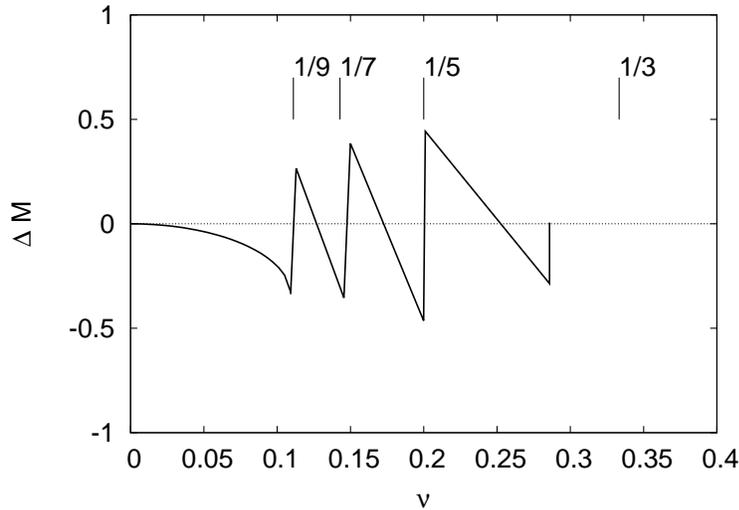}
\caption{Schematic diagram of the change in magnetization $\Delta M$ on melting along the
melting curve shown in Fig.~\protect\ref{fig:1p100book}. Redrawn after
Ref.~\protect\citelow{Lea:91}.
}
\label{fig:2p100book}
\end{figure}

Lea \emph{et al.} \cite{Lea:91} gave the thermodynamics of the melting of a 2D
electron assembly in a magnetic field. Their theoretical study was prompted by
the experiment of Buhmann \emph{et al.} \cite{Buhmann:91} on the luminescence
spectrum of a heterojunction. This led these authors to sketch the melting curve
of the Wigner solid as a function of the Landau level filling factor $\nu$. This
is defined in terms of the (areal) electron density $n$ and the magnetic field
strength by
\begin{equation}
\nu = n h c/e B.
\end{equation}
The schematic diagram sketched by Buhmann \emph{et al.} \cite{Buhmann:91} has
been interpreted thermodynamically by Lea \emph{et al.} \cite{Lea:91} and the
main results are summarized in Figs.~\ref{fig:1p100book} and \ref{fig:2p100book}.

Lea \emph{et al.} \cite{Lea:91} noted first that the thermodynamics of an
electron crystal to electron liquid first-order melting transition leads for the
melting temperature $T_m$ as a function of magnetic field to the result, at
constant area $\Omega$,
\begin{equation}
\left( \frac{\partial T_m}{\partial B} \right)_\Omega = - \frac{\Delta M}{\Delta
S} .
\end{equation}
Denoting the crystal phase by subscript $C$ and the liquid by $L$, then $\Delta
M = M_L - M_C$ is the change of magnetization on melting, while $\Delta S = S_L
- S_C$ is the corresponding change in entropy. This is then cast by Lea \emph{et
al.} \cite{Lea:91} in terms of the Landau filling factor $\nu$ as
\begin{equation}
\left( \frac{\partial T_m}{\partial \nu} \right)_\Omega = \frac{B}{\nu} \frac{\Delta M}{\Delta
S} .
\end{equation}

The phase diagram in Fig.~\ref{fig:1p100book} showing the equilibrium melting
curve between the Wigner electron solid and the Laughlin electron liquid shows
four crystal phases $C_1$ to $C_4$, and the relevant liquid phase at
$\nu=\frac{1}{9}$, $\frac{1}{7}$, and $\frac{1}{5}$. The reader interested in
finer details should consult the work of Lea \emph{et al.} \cite{Lea:91}.

\subsection{Anyon magnetism in the Laughlin liquid near the melting curve of the
Wigner electron solid}

After their thermodynamic understanding of the above melting curve, Lea \emph{et
al.} \cite{Lea:91} sought macroscopic understanding of especially the change in
magnetization $\Delta M$ across the melting curve. It then became clear that
most of the remarkable features of the melting curve must reside in the
magnetism of the Laughlin liquid near the melting temperature. And since
knowledge of the magnetism of the Wigner electron solid pointed to weak
antiferromagnetism, $\Delta M$ must come from the 2D electron liquid. This led
to the treatment of anyon magnetism, due to the fact that in 2D the electron
liquid obeyed anyon \cite{Leinaas:77,Wilczek:82,Wilczek:82a} rather than Fermi
statistics.

To complement the above thermodynamics, the magnetism of the Laughlin liquid was
studied by Lea \emph{et al.} \cite{Lea:91} using the anyon model (see also
Refs.~\citeonline{Pellegrino:07,Pellegrino:08} and references therein on fractional
statistics). Assuming this dominates $\Delta M$ in the thermodynamics, since the
Wigner solid exhibits only weak cooperative magnetism, Lea \emph{et al.}
\cite{Lea:91} drew a schematic diagram (Fig.~\ref{fig:2p100book}) of the change
in magnetization on melting along the melting curve shown in
Fig.~\ref{fig:1p100book}. As these authors stressed, the field dependence of
$\Delta M$ is very reminiscent of the de~Haas-van~Alphen effect
\cite{Jones:73-1} at integral values of $\nu$.

To conclude this brief discussion, we should note that Wu \emph{et al.}
\cite{Wu:98} (see also Ref.~\citeonline{March:00a}) have given a detailed
theoretical treatment of the thermal activation of quasiparticles and the
thermodynamic observables in fractional quantum Hall (FQH) liquids. But their
final sentence is relevant to the work of Lea \emph{et al.} \cite{Lea:91} under
discussion here. This sentence reads: `In particular, it is more desirable that
these theoretical predictions \cite{Wu:98} would be put to experimental tests,
if the tremendous difficulties in measuring thermodynamic quantities of a thin
layer of electron gas could be overcome someday.' It is therefore the more
remarkable that existing experiments on GaAS/AlGaAs heterojunctions in magnetic
fields ${\bf B}$ perpendicular to the two-dimensional electron assembly can
provide crucial insight into some of the thermodynamic quantities characterizing
the FQH liquid studied theoretically by Wu \emph{et al.} \cite{Wu:98}.

\section{Summary and future directions}
\label{sec:conclusions}

Though different in emphasis from the approach adopted in the present review, it
is relevant here to refer to a body of work on bonding and information theory.
The interested reader is recommended to start by using especially the early
references in the article by Nalewajski \cite{Nalewajski:06}. In this particular
article, the focus is on H$_2$, a comparison being made between the valence bond
treatment and communication theories.  In the above context, further insight has
been obtained into the work of Hirshfeld \cite{Hirshfeld:77}. Relevant
references can also be found in Parr \emph{et al.} \cite{Parr:05}. In relation
to emergent properties, the correction between quantum entanglement and
correlation in many-body assemblies is considered in the work of Campos~Venuti
\emph{et al.} \cite{CamposVenuti:05,CamposVenuti:06}.

To conclude, we want to refer again to the relevance of quantum information
theory in many-body theory. In particular, we recommend the reader to consult
the discussion of Schollw\"ock \cite{Schollwoeck:05} (see especially section~C).
Schollw\"ock stresses in particular, in quantum chemistry applications, the
importance of the entanglement entropy. The work of Legeza and S\'olyom
\cite{Legeza:03} is another good starting point for the interested reader.
Finally, for the reader needing further details on quantum information, we
recommend study of the book by Nielsen and Chuang \cite{Nielsen:00}.

\section*{Acknowledgements}

NHM wishes to acknowledge that his contribution to this review was greatly aided
by the invitation of Professor \'A. Nagy to deliver a plenary lecture at a week
long Workshop in the University of Debrecen, Hungary, covering both information
theory and emergente properties. Stimulating discussions with her and other
participants to the Workshop are gratefully acknowledged. NHM is also indebted
to Professors C. Amovilli and A. Rubio, and to Dr A. Akbari for much
collaboration and valuable discussions. Much of NHM's contribution was made
during a visit to the University of Catania, Italy, and he thanks Professors R.
Pucci and G. G. N. Angilella for their kind hospitality.

\appendix{Relation between three forms of quantum information entropy for a
whole class of two-electron spin-compensated models}

Holas \emph{et al.} \cite{Holas:03} have solved for the 1DM a class of models
for two-electron spin-compensated atomic ions. Below, we first summarize the
essence of their results and then use these to set out a route for relating the
three forms of quantum information entropy set out in Eqs.~(\ref{eq:Shannon}),
associated with the name of Shannon, plus Eq.~(\ref{eq:Jaynes}), defining Jaynes
entropy.

For the case when we choose the constants in Eqs.~(\ref{eq:Shannon}) as $-1$, to
accord for example with Horny\'ak and Nagy \cite{Hornyak:07} (see also
Ref.~\citeonline{Nagy:13}), who give an Uncertainty Principle inequality
satisfied by the Shannon forms sum $S_\rho + S_n$, namely
\begin{equation}
S_\rho + S_n \leq 1 + \ln\pi .
\end{equation}
It is proposed below how one can, in the future, numerically test the utility
of this inequality for the above class of two-electron artificial atoms. Also, the
relation between the 1DM and the electron density $\rho(\br)$ will be given for
the Moshinsky model in which both external potential $V_{\mathrm{ext}}$ and interaction
$u(r_{12})$ have harmonic forms. 

The essence of Holas \emph{et al.} \cite{Holas:03} was to separate the center of
mass ($cm$) part from the
relative motion ($rm$) part of the ground-state wave function as
\begin{equation}
\Psi(\br_1 , \br_2 ) = \Psi_{cm} \left( \frac{\br_1 + \br_2}{2} \right)
\Psi_{rm} (|\br_1 - \br_2 |).
\end{equation}
The latter rm part solves the one-body Schr\"odinger equation with effective
potential $V_{\mathrm{eff}} (\br) = V_{\mathrm{ext}} (\br) + \mathrm{const} \times u(r_{12} )$. The
analytical form of the cm part is \cite{Akbari:13}

\begin{equation}
\Psi_{cm} (R) = \exp\left( -\frac{1}{2} \frac{R^2}{a^2_{cm}} \right) ,
\end{equation}
where $a_{cm} = (k/2m\omega_0)^{1/2}$, and $\omega_0^2 = k/m$. Writing the $rm$
wave-function as $\Psi_{rm} (r) = \mathrm{const} \cdot \psi_{rm} (r) /r$, the one-body
Schr\"odinger equation below results:
\begin{equation}
\left[ -\frac{\hbar^2}{2m_{rm}} \left( \frac{d}{dr} \right)^2 
+\frac{1}{2} m_{rm} \omega_0^2 r^2 + u(r) \right] \psi_{rm} (r) =
E_{rm} (r) \psi_{rm} (r).
\end{equation}

Holas \emph{et al.} \cite{Holas:03} obtained $\gamma(\br,\br^\prime)$ exactly by
quadrature on $\Psi_{cm}$ and $\psi_{rm}$. However, the form of $\gamma$ is
somewhat complex, so below we limit ourselves to reporting
$\left.\gamma(\br,\br^\prime) \right|_{\br=\br^\prime} = \rho(\br)$ explicitly
for general $u(r)$ as
\begin{equation}
\rho(r) = \frac{8}{\sqrt{\pi}} e^{-r^2 /a_{cm}^2}
\int_0^\infty dy\, y^2 e^{-y^2/4} [\Psi_{rm} (a_{cm} y)]^2
\frac{\sinh (ry/a_{cm})}{ry/a_{cm}} .
\label{eq:sinh}
\end{equation}

One can solve Eq.~(\ref{eq:sinh}) by Fourier transform (FT) to obtain $\Psi_{rm}$ in
terms of the FT of $\rho(\br)$: usually called the atomic scattering factor
$f(\bk)$. But $\gamma(\br^\prime,\br)$, the exact 1DM, is also known as characterized by
$\Psi_{rm}$. Therefore, for this class of models, the three information
entropies can all be related to the diagonal density $\rho(\br)$, most elegantly
in terms of the scattering factor $f(\bk)$.

\begin{figure}[t]
\centering
\includegraphics[height=0.8\columnwidth,angle=-90]{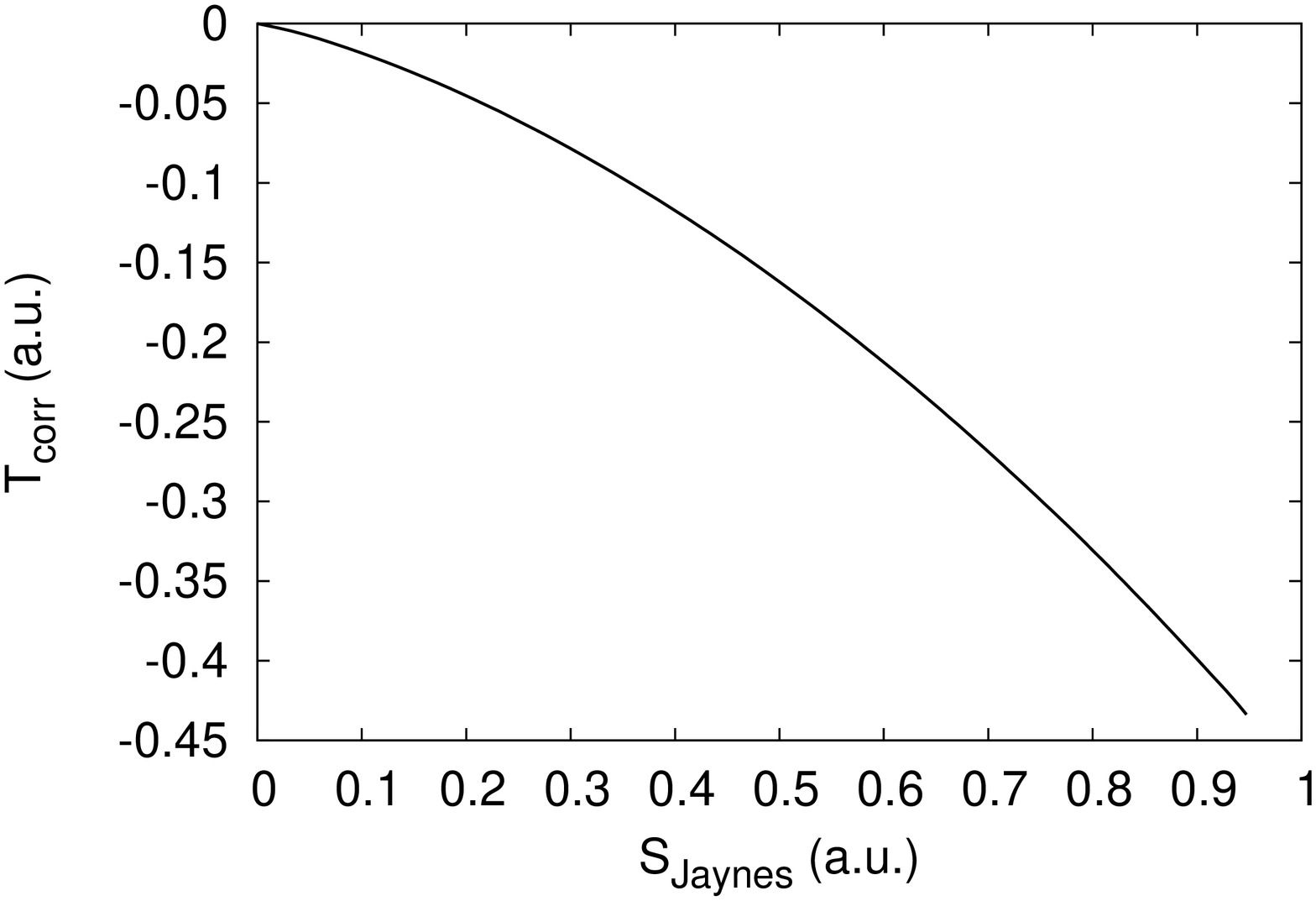}
\caption{Correlation kinetic energy against Jaynes entropy, both in atomic
units, for the Moshinsky atom. Redrawn after
Ref.~\protect\citeonline{Amovilli:04b}.}
\label{fig:Amovilli4}
\end{figure}

For the Moshinsky atom, Holas \emph{et al.} \cite{Holas:03} obtained the
relation between the 1DM $\gamma$ and the ground-state density $\rho(r)$ as
\begin{equation}
\frac{\gamma(\br_1 , \br_1^\prime )}{\gamma(\br_0 , \br_0 )}
= \left( \frac{\rho(\bar{r})}{\rho(r_0)} \right)^{\alpha^2 /(2\alpha-1)} ,
\label{eq:gammaalpha}
\end{equation}
where $\br_0 = \frac{1}{2} (\br_1+\br_1^\prime)$, $\bar{r} = [\frac{1}{2} ( r_1^2 +
r_1^{\prime 2} )]^{1/2}$, and the `interaction strength' parameter $\alpha$ is
defined as
\begin{equation}
\alpha = \frac{1}{2} ( 1 + \sqrt{1+2K} ) ,
\end{equation}
which can in fact be related to the density $\rho(r=0)$ by
\begin{equation}
\alpha^{-1} = 2 - \pi [\rho(r=0)/2]^{2/3} .
\end{equation}
Eq.~(\ref{eq:gammaalpha}) above agrees with the earlier result given by March
\cite{March:02b} apart from the normalization factor.
Holas \emph{et al.} \cite{Holas:03} also showed that the kinetic energy density
for the Moshinsky atom is related to the ground-state density $\rho(r)$ as
\begin{equation}
t(r) = \frac{1}{2} \rho(r) \left( \frac{3}{2} \frac{(\alpha-1)^2}{\alpha} -
\frac{2\alpha-1}{\alpha} \ln \frac{\rho(r)}{\rho(0)} \right).
\end{equation}
Fig.~\ref{fig:Amovilli4} shows a plot of the correlation kinetic energy against
Jaynes entropy for the Moshinsky atom \cite{Amovilli:04b}.

\section*{References}

\bibliographystyle{ijmpb}
\bibliography{a,b,c,d,e,f,g,h,i,j,k,l,m,n,o,p,q,r,s,t,u,v,w,x,y,z,zzproceedings,Angilella}

\end{document}